\documentclass[aps,prl,twocolumn]{revtex4}
\usepackage{graphicx}
\usepackage{amsmath}
\usepackage{epsfig}

\begin{document}
\def\be{\begin{equation}}
\def\ee{\end{equation}}
\def\bearr{\begin{eqnarray}}
\def\eearr{\end{eqnarray}}
\def\tc{$T_c~$}
\def\tcl{$T_c^{1*}~$}
\def\c2{ CuO$_2~$}
\def\ruo{ RuO$_2~$}
\def\lsco{LSCO~}
\def\bi{bI-2201~}
\def\tl{Tl-2201~}
\def\hg{Hg-1201~}
\def\sro{$Sr_2 Ru O_4$~}
\def\rc{$RuSr_2Gd Cu_2 O_8$~}
\def\mgb{$MgB_2$~}
\def\pz{$p_z$~}
\def\ppi{$p\pi$~}
\def\sqo{$S(q,\omega)$~}
\def\tperp{$t_{\perp}$~}
\def\cob{$\rm{CoO_2}$~}
\def\nxcob{$\rm{Na_x CoO_2.yH_2O}$~}
\def\ncob{$\rm{Na_{0.5} CoO_2}$~}
\def\half{$\frac{1}{2}$~}
\def\nycob{$A_xCoO_{2+\delta}$~}
\def\tj{$\rm{t-J}$~}
\def\hlf{$\frac{1}{2}$~}

\title{Resonating Valence Bond States in 2 and 3D\\
-- Brief History and Recent Examples}

\author{G. Baskaran}
\affiliation{Institute of Mathematical Sciences\\
Chennai 600 113, India}
\date{\today}

\begin{abstract}
Resonating Valence Bond states are quantum spin liquids, having low
energy spin-\hlf (spinon) or spin-1 excitations. Although spins are
`disordered', they posses subtle topological orders and some times
chiral orders. RVB states are easily appreciated and seem natural in
the quantum fluctuation dominated 1D world. In 2 and 3D, competing
orders such as antiferromagnetism, charge order or even
superconductivity often hide an underlying robust quantum spin
liquid state. Introduction of additional spin interactions or doping
of delocalized charges, or finite temperatures, could frustrate the
long range magnetic order and reveal a robust RVB state. To this
extent they are natural in 2D and above. We present a brief history
of insulating RVB states. Then we summarise our own recent theory of
RVB states for 2 and 3D systems, including some newly synthesised
ones: i) boron doped diamond, ii) \nxcob, iii) quasi 2D organic
conductors and iv) a 2D graphene sheet.

\end{abstract}

\maketitle

\section*{\textsc{I. Introduction}}

Resonating valence bond states occupy a special position among
quantum states and phases in condensed matter physics. They became
popular and important, as a seat for high temperature
superconductivity\cite{bednorz, pwascience} in cuprates.
Subsequently, their novel quantum properties, such as quantum number
fractionization, topological order and a deep connection to gauge
theories, have also received a well deserving attention in the past
two decades. In this article, we give a brief history of RVB theory,
focussing on the insulating states, followed by a short introduction
to RVB states. Then we provide some examples for RVB states from
real systems in 2 and 3D, based on our own results and new
understanding that have come in the last few years: i) boron doped
diamond, ii) \nxcob, iii) quasi 2D organic conductors and iv) a 2D
graphene sheet.

\section*{\textsc{II. A Brief History}}
The idea of resonating valence bond (RVB) arose in the description
of quantum mechanical resonance of covalent bonds in unsaturated
p-$\pi$bonded organic molecules such as benzene. It was
soon generalised to 2D graphite and metals by Pauling\cite{pauling}.
The overwhelming phenomenological success of semi empirical results
and reasonings of Pauling, apparently even questioned the need for
notions such as Fermi surface in describing a metal. It was at this
point, in 1973, Anderson \cite{pwa73} became enthusiastic about the
idea of RVB in a Mott insulator, while he remained silent about
metals. He pointed out that the idea of RVB could be really relevant
to family of spin-\hlf Mott insulators, where an expected long range
antiferromagnetic order was often absent. Anderson attributed this
to enhancement of quantum fluctuations created by frustrated spin
interactions and lower dimensionality. He elaborated this by an
analysis of a 2D triangular lattice of spin-half Heisenberg
antiferromagnet. His variational study showed that this system could
very well have a quantum spin liquid ground state, a short range RVB
state.

Very few in condensed matter community paid attention to this
proposal; exceptions were Fazekas\cite{pwaFazekas}, Klein, Shastry,
Sutherland, Caspers, some Japanese experimental groups (Hirakawa,
Yamada and possibly others) and to some extent myself. I was
familiar with RVB ideas, partly through Fazekas in the early 80's,
at ICTP, Trieste, a wonderful meeting ground of so called third and
first world. (A first rate condensed matter theory activity, that
continues now, was being nurtured by the efforts of Stig Lundquist,
Norman March, Paul Butcher, Eli Burstein, Abdus Salam and faculty
like Mario Tosi, Erio Tosatti, Michele Parrinello, Roberto Car and
Yu Lu and others. People like Phil Anderson, Bob Schrieffer took great
interest in ICTP activities and frequented the center. My
association with Anderson was made possible by ICTP and Erio
Tosatti).

In the early 1986, just before Bednorz-Muller's discovery was
published, I was intrigued by the question of phase relations among
valence bond configurations in a short range RVB wave function, and
what it really meant in a magnetic insulator. When Anderson
responded\cite{pwascience} to Bednorz-Muller's discovery of high Tc
superconductivity, with his RVB proposal, I was well prepared and
was quick to appreciate it. That is how I
joined Anderson in his second RVB journey, as a partner. A 20 year
long journey is still continuing. Anderson's Science
paper\cite{pwascience} and our
collaborative works\cite{bza, gauge, abhz}, interestingly, done
during the very first year of this journey (1987-88), continue to
light the path.

On another front, a non-trivial solution to 1D antiferromagnetic
Heisenberg chain by Bethe\cite{BAnsatz} was drawing more and more
attention; its mathematics was formidable and revealed surprises
such as, absence of long range antiferromagnetic order, even at T =
0 and presence of gapless topological (domain wall) spin half
excitations\cite{fadeevTakhtajan}, which was later named\cite{abhz}
spinon. It is in this background, before Anderson proposed his RVB
theory in 1973, Majumdar and Ghosh discovered a model\cite{MGmodel}
in 1969, a slight variant of standard Heisenberg chain, which
exhibited a strikingly simple many body ground state - a valence
bond solid. Valence bond resonance was completely absent.

Even these developments did not suggest an RVB description of the
complicated Bethe ansatz ground state. Shastry,
Sutherland\cite{shastrySutherland}, Klein\cite{klein} and others
went for higher dimensional generalisation of Majumdar-Ghosh model
and valence bond solid phases. Interestingly, recent
study\cite{klein2} of Klein models have given a rich possibility of
quantum liquid of valence bonds, at some special lattices and for
some choice of parameters. The fertile modern materials science has
offered a compound \cite{SScompound}, SrCu$_2$(BO$_3$)$_2$, where
the Shastry-Sutherland model is indeed realized.

The idea of resonating valence bond was in the hands of quantum
chemists for a long time, mostly studying the p-$\pi$ bonded organic
systems. The richness of this novel quantum phase was yet to be
unravelled. According to Anderson, when superconductivity in oxides
such as LiTi$_2$O$_4$ and BaBi$_{1-x}$Pb$_x$O$_4$ were
discovered\cite{BaBiO} thoughts of RVB crossed his subconscious
mind. Like many cases in physics, a key
experimental result was necessary to open the doors and revive and
flourish an old and fertile idea such as resonating valence bond.
Discovery of high Tc superconductivity in an unexpected oxide
La$_{1-x}$Ba$_x$CuO$_4$ by Bednorz and Muller catalysed a revolution
in the RVB theory front. As mentioned earlier, it was Anderson, who
was, alert and sensitive to the new oxide and the challenge from
experiments. He proposed the RVB mechanism of superconductivity. His
collaborators gave flesh to his proposal and offered new insights.
Two key many body approaches were developed: i) RVB mean field
theory\cite{bza} (BZA theory) and ii) a gauge
theory\cite{gauge,gauge2} (BA theory), to go beyond RVB mean field
theory. The currently popular experimental phase diagram for
cuprates was part of the RVB theory conjecture\cite{abhz}, before
the experimental phase diagram emerged.

The RVB character of the ground state of 1D Heisenberg Chain (Bethe
Ansatz wave function) was well recognised in the BZA paper. A new
impetus was given by Haldane and Shastry, who showed that the
Gutzwiller projected BZA mean field solution in 1D is indeed the
ground state of a non-trivial 1D Heisenberg model, which has become
the celebrated Haldane-Shastry model\cite{HaldaneShastry}.

The BZA and BA theory showed a deep connection between RVB states
and gauge theories; quantum number fractionization came out as a
rather natural possibility. A gauge structure and dynamically
generated gauge fields in a quantum spin problem was rather
unexpected and opened a new field of activity. Using these insights
a new mean field solution by Affleck and
Marston\cite{AffleckMarston} and a chiral spin liquid wave function
by Kalmayer and Laughlin\cite{KalmayerLaughlin} appeared in the
scene. They contained a nonzero ground state condensates of the BA
gauge fields. Gauge field condensation, flux tube attachment and a
consequent statistics transmutation eventually lead to Laughlin's
proposal\cite{LaughlinAnyonSC} of anyon superconductivity in 2
dimensions. Wen, Wilczek and Zee\cite{wwz} made a key identification
of the `magnetic flux' of RVB gauge field with spin chirality, ${\bf
S }_i\cdot({\bf S}_j \times{\bf S}_k$). Large N theories\cite{readN}
that followed Affleck-Marston's work studied valence bond solid
phases.

Kotliar's d-wave RVB meanfield solution\cite{kotliar}, based on
slave boson approach, adapted to t-J model by Zou and
Anderson\cite{zouAnderson}, explained the d-wave symmetry of cuprate
superconductors successfully; Fukuyama\cite{fukuyamaRVB} school and
others did extensive study on this front.

Kivelson, Rokhsar and Sethna\cite{KRS} used short range RVB wave
function to study cuprates and introduced the notion of `holon', a
topological excitation for charge. Other authors, including
Sutherland\cite{sutherlandRVB} and Reed and
Chakraborty\cite{ReedBulbul} pursued the study of short range RVB
states and the nature of spinon and holon excitations. Quantum dimer
models introduced by Kivelson and Rokhsar\cite{KR}, to understand
short range RVB states have brought out novel topological ground
state degeneracies, and some non-trivial gapless spin liquid phases;
this has been developed further by Sondhi and
Mesmer\cite{SondhiRVB}, and others. The idea of topological
degeneracy, that also unifies RVB states with fractional quantum
Hall states, has been elevated to an interesting notion of `quantum
order' by Wen\cite{Qorder} and several insights have been offered.

Statistics other than fermion and boson were suggested to be
possible in a 2D world by Leinaas, Myrlheim, Wilczek and
Zee\cite{LlinasWilczekZee}. RVB phases and quantum Hall systems
became play grounds for particles with non standard statistics.
Dzyaloshinski, Wiegman and Polyakov\cite{Hopf}, suggested
interesting statistics transmutation properties for spinons, through
a topological Hopf term, for the 2D spin-\hlf Heisenberg
antiferromagnet. Some attempts\cite{haldaneBF, reedBF} to organize
the sum of single spin Berry phase terms did not lead to the
anticipated Hopf term. However, in a recent work\cite{GBskyrmion}
the present author has shown that a proper summation of the Berry
phase terms leads effectively to a statistics transmutation. The
Berry phase does get organized and behave like a non-trivial
topological term; however, it does not have a local continuum Hopf
like analytic form.

Affleck, Zou, Anderson and Hsu\cite{SU2Anderson} and also Fradkin,
Moreo, and Dagotto\cite{SU2Fradkin} found a SU(2) description of the
BA theory. However, it was realized soon that owing to the limited
dimensionality of Hilbert space of our spin system, only a subgroup,
the center Z$_2$ of the SU(2) group is really necessary to describe
the thermodynamic phases and dynamics, rather than the full SU(2) or
U(1) group. This was nicely shown by an identity due to
Marston\cite{marstonChernSimons}, which showed how the dynamically
generated RVB flux gets restricted to integer or half integer flux
quanta rather than an arbitrary value. That is at the level of a
classical action, SU(2) or U(1) fields exists formally. However, the
quantum dynamics chooses only a limited set of the field degrees of
freedom. Marston\cite{marstonChernSimons} incorporated the quantum
kinematic restriction through a Chern-Simons term in the action, by
hand. What is important is that within the subspace of zero and half
flux, the Chern-Simons term retains the PT symmetry. This was soon
taken further and a Z$_2$ gauge theory of spin system was formulated
by Tosatti, Yu Lu and the present author\cite{gbYuLuTosatti}. In
another work, using a similar identity due to Wen, Wilczek and
Zee\cite{wwz}, the present author\cite{gbChiral} reduced the famous
triangular lattice problem to a Z$_2$ gauge theory. Zou in an
insightful paper\cite{zou1} discussed how Chern Simon terms could
arise as a quantum anomaly in an SU(2) gauge theory for spin-\hlf
Heisenberg antiferromagnet in 2D. Wen\cite{wenZ2} and Read and
Sachdev\cite{readZ2} also developed the Z$_2$ gauge theory ideas and
connection to topological degeneracies etc.

Systematic way of going beyond BZA theory for insulating and
conducting spin systems using Gutzwiller approximation has become
very useful for quantitative progress, in the hands of Gros, Zhang,
Rice\cite{newRVB, ogataGW}, Ogata, Shiba,  and recently
Randeria, Trivedi, Paramekanti\cite{nandini},
Muthukumar\cite{muthGW} and others. The BA gauge theory, on the
other hand has been very useful in giving new qualitative insights;
its full potential as a quantitative tool has not been realized,
in spite of notable efforts\cite{nagaosa} by Ioffe, Larkin, Nakamura,
Matsui, Patrick Lee, Nagaosa, Wen, Dung-Hai Lee, Ng and recently
Tesanovic, Franz and others.

Hsu\cite{hsuAFM}, showed that the antiferromagnetic order existing
in the ground state of 2D Heisenberg model on the square lattice can
be viewed as a spinon density wave in an underlying quantum spin
liquid. A `bosonic' variational RVB wave function (similar to
Gutzwiller projection of Arovas Auerbach's Schwinger
boson\cite{arovasAuerbach} type wave function) introduced by Ducout,
Liang and Anderson\cite{doucotLiangPWA} exhibited a spontaneous
antiferromagnetic order in the ground state for a range of
variational parameter. Outside this range the spin correlation
function decayed exponentially. However, the energy expectation
value changed very little with the variational parameter, even
though sub lattice magnetisation changed substantially from zero to a
large value. This analysis substantiated the fact that long range
antiferromagnetic order is a minor modification in an otherwise
robust spin liquid state. Some of these ideas have been summarised
by a principle of valence bond amplitude maximisation
(VBAM)\cite{GBvbam} by the present author.

In the recent past, quantum number fractionization and spinon
deconfinement has been studied by Fisher, Balents, Nayak, Senthil,
Viswanath, Sachdev and collaborators\cite{SenthilFisher}. A Z$_2$ gauge
symmetry has been very prominent in the discussion. Possibilities of
classifying RVB states into Z$_2$, U(1) and non-abelian spin liquids
have been discussed.

A recent work by the present author shows\cite{GBskyrmion} a
surprising result that quantum number fractionization occurs, above
a finite energy gap, even in the ordered Heisenberg antiferromagnet
in 2D ! That is, in addition to gapless spin wave excitations we
have deconfined, freely propagating spinons above a finite energy
gap. I showed that a (scale free) finite energy quantum skyrmions is
made of two unbound `chiral spinons'. Chiral spinons carry non
vanishing condensed RVB magnetic flux or chiral density
${\bf S }_i\cdot({\bf S}_j \times{\bf S}_k$)
distributed specially in a broad fashion. This result confirms and
sharpens an early conjectured connection of meron with spinon by
John, Doucot, Liang, Anderson and Baskaran\cite{JohnDoucotMeron,
johnBirceau}. It will be interesting to make connection of our
formally exact result with recent works of Weng, Ng,
Muthukumar\cite{muthuSpinon, muthuWeng} and collaborators d
on spinons.

An elegant construction by Affleck, Kennedy, Lieb and
Tasaki\cite{AKLT} has given models with valence bond like ground
states with higher spins and higher dimensions. In the process it
has given a new meaning to the Haldane gap phenomenon.

On numerical front, RVB wave functions have been analysed by several
authors for frustrated and non-frustrated spin systems in great
detail. Highly frustrated spin systems such as Kagome lattice has
given some surprises\cite{Luhlier}.

RVB excitations, because of their topological and `abelian' or
`non-abelian anyon' character, arising from topological degeneracy
in the ground state, could have a special immunity against
decoherence. They also have fascinating quantum entanglement and
braiding properties. Consequently, they have been considered as
serious q-bit candidates in quantum computers by Kitaev\cite{kitaev}
and others. RVB theory and fractional quantizaed Hall effects have
indirectly given a new impetus to theoretical studies in quantum
computers, with envisaged experimental potential\cite{dasSarmaQbit}.

This is a brief history of RVB theory, without going to the
fascinating superconductivity or antiferromagnetism aspects.

\section*{\textsc{III. RVB wave function, Topological Degeneracy and Excitations}}

Above $T_N$ thermal fluctuations destroy long range
antiferromagnetic order in quantum spin systems. At very high
temperature the thermal state is a structureless `classical'
paramagnetic phase. What is the state we reach, when we destroy long
range antiferromagnetic order, by frustrating it through additional
interactions, at T = 0 or k$_B$T $ \langle \langle$ J ?
The `spin crystal' quantum
melts and we get a quantum spin liquid. In this quantum spin liquid,
the antiferromagnetic order decays in a power law or exponential
fashion. This phase, where spins are seemingly disordered, have some
special quantum coherence properties, which is what makes it a
resonating valence bond state. This paramagnetic state has a special
pair coherence among spins and also topological degeneracy.

The special pair coherence has a natural and suggestive
representation as a general RVB state, written down first by
Anderson\cite{pwascience}. It turns out that this state has a rather
natural representation, not in the standard $S^z$ basis, but in
terms of underlying electron operators c's that makeup a spin half
moment:

\be |RVB;\phi\rangle \equiv P_G (\sum_{ij} \phi_{ij}
b^\dagger_{ij})^{N\over2} |0\rangle \ee

where, $b^\dagger_{ij} \equiv {1\over{\sqrt
2}}(c^\dagger_{i\uparrow}c^\dagger_{j\downarrow} -
c^\dagger_{i\downarrow}c^\dagger_{j\uparrow})$.
P$_G\equiv\prod_i(1-n_{i\uparrow}n_{i\downarrow}$), is the
Gutzwiller projection, which ensures that the effective low energy
electron occupancy of any site in a Mott insulator is one.
Therefore, total number of electrons N in equation (1) is the same as
the number sites. The pair function $\phi_{ij}$ characterises the
RVB state. The RVB wave function (equation 1) is identical to,
except for the Gutzwiller projection, an N particle projected BCS
wave function, with $\phi_{ij}$ playing the role of a Cooper pair
function. This is what made Andersons proposal of a (Mott)
insulating RVB state becoming a superconductor, on moving away from
half filling (doping), so natural and appealing.

In a 2D square lattice, the standard short range RVB corresponds to
$\phi_{ij}$ non zero only for nearest neighbour sites and a special
relation between signs of $\phi$'s between neighbouring bonds, so as
to satisfy Marshall sign convention. In general, for various choices
of $\phi_{ij}$ we get i) the BZA state with a pseudo fermi surface
ii) Affleck-Marston $\pi$-flux state with nodal excitations iii)
gapful Kalmayer Laughlin's chiral spin liquid state iv)
antiferromagnetically ordered state v) d, d+id and s-wave
superconducting state iv) ground state of Haldane Shastry
Hamiltonian in 1D v) states with charge and spin stripe correlations
etc. The physically motivated Gutzwiller projection does wonders -
it enhances antiferromagnetic correlations and even introduces
strong chiral correlations, in the process of reducing the double
occupancy fluctuations.

Anderson's RVB wave function and the corresponding Hilbert space of
states for strongly correlated electron systems is as basic and
similar to
Laughlin wave function and the corresponding Hilbert space of states
for quantum Hall physics. It is very different from effectively
slater determinant type `Fermi liquid Hilbert space'.

Kivelson and Rokhsar introduced and studied a quantum dimer model on
a square lattice, with a view to understand short range RVB physics.
This non-trivial model has given many insights. For example, the
idea of topological degeneracy in the ground state was manifest. If
we consider a square lattice with periodic boundary condition (torus
of genus g) all dimer coverings break into 4$^g$ distinct classes,
such that they are super selected with respect to local moves of the
valence bonds. In the RVB mean field theory the topological
degeneracy appeared as a (PT symmetric) half flux quanta of magnetic
flux of RVB gauge field, that can be be thread through various holes
in the torus of genus g. In this sense there is a close connection
of this topological degeneracy with corresponding one in fractional
quantised Hall states, giving the possibility of non-abelian
character to spinons, in some RVB states.

\begin{figure}
\includegraphics*[width=6.0cm]{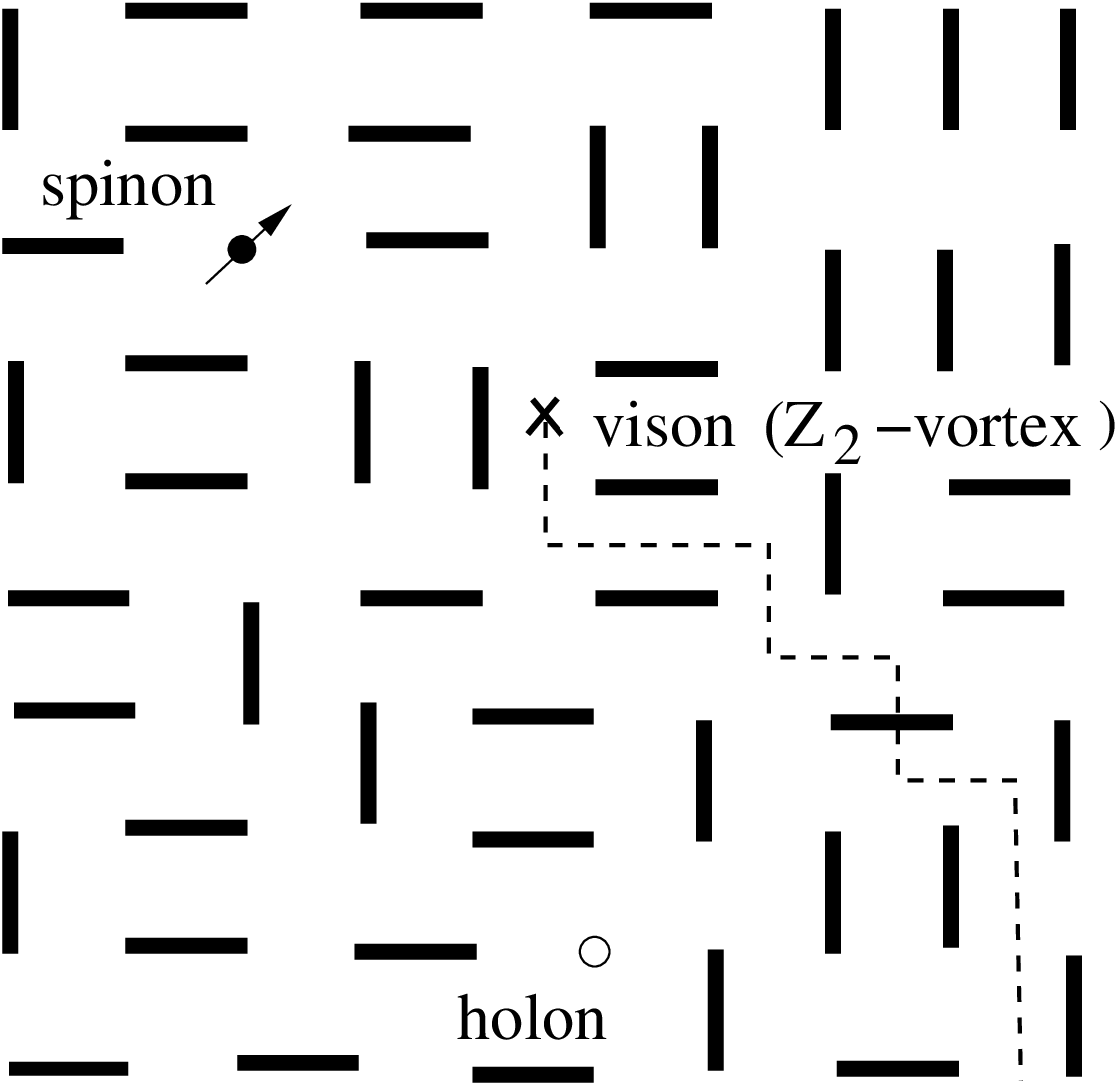}
\caption{\label{fig1} Schematic pictures of topological excitations,
i) spinon, ii) holon and iii) a Z$_2$ vortex (vison). The valence bonds
in the background should resonate in the readers mind.}
\end{figure}

The valence bond character of the RVB wave function suggests
presence of certain type of topological excitations (figure 1) : i)
spinon: an unpaired spin in the background of resonating singlets
and ii) holon or doublon: an empty site or a doubly occupied site in
the background of resonating singlets and iii) Z$_2$ vortices,
carrying a $\pi$-flux, dubbed as `vison' in recent
works\cite{SenthilFisher}.

A simple way to imagine a spinon is to freeze a singlet bond and
convert it into a triplet and localise one up spin at a given site
and move the other upspin to the boundary. What we get is a
localised spinon, an unpaired upspin in the background of resonating
singlets. For some RVB systems such an unpaired spin may become
freely propagating spinon excitation; examples are 1D Heisenberg
chain, 2D chiral spin liquid, BZA phase with a spinon pseudo fermi
surface, Affleck-Marston phase with nodal spinons etc. If
this be the case we have quantum number fractionization and spinon
deconfinement. In some cases two spinons may be bound and we may get
a spin-1 low energy excitation branch. In the case of short range
RVB system in 2D and 3D (without any chiral symmetry breaking) we
expect spinons to be confined; however the spin-1 branch is a well
defined excitation of the underlying quantum spin liquid. The
Z$_2$ vortices are best understood as a `local defect' in the
Marshal sign convention, which carries a phase string. The phase
string excitation have bee studied in detail by Weng and
collaborators\cite{muthuWeng}. Energetic considerations and
some deep issues related to confinement may force either a spinon or
holon to be bound to a Z$_2$ excitation.

In the valence bond basis, creating a spinon at a site is a
complicated non local operation, as outlined. The nonlocal and
global rearrangement needed to create an isolated spinon or holon
make them topological excitations. BZA theory give a simple and
straightforward method to construct spinons and holons. In this
approach it is done by a local operation of creating in the mean
field RVB state (that lives in an enlarged Hilbert space) a
particle-hole pair excitation followed by a Gutzwiller projection:
\be \zeta^\dagger_{i\sigma} \zeta^\dagger_{j\sigma'} |RVB\rangle
\equiv P_G c^\dagger_{i\sigma} c^{}_{j-\sigma'} |mRVB\rangle \ee
where $|mRVB\rangle = (\sum_{ij} \phi_{ij}
b^\dagger_{ij})^{N\over2}|0\rangle $ is the unprojected RVB mean
field solution. Here $\zeta_{i\sigma}$ is a spinon operator. We note
that the order of operation is important: Gutzwiller projection
should be done after the creation of particle-hole pair. We can
construct holon or doublon in a similar fashion.

Using the above construction, non-trivial excitations such as the
spinon of the Haldane Shastry model in 1D and the Kalmayer-Laughlin
model in 2d can be easily constructed. Somen Bhattacharjee's
pfaffian representation\cite{somen} of RVB wave function may be
useful in the study of RVB states; it remains to be explored.

\section*{\textsc{IV. New Examples of 2 and 3D RVB States}}

In what follows we will summarise our recent results for certain
non-cuprate systems, for which we have suggested RVB phases as the
suitable reference vacuum state. We will not go to details of the
theory, but will only outline the physics behind. The systems are:
i) quasi 2D organic conductors ii) boron doped diamond, iii) \nxcob
and iv) a 2D graphene sheet.

\begin{figure}
\includegraphics*[width=8.0cm]{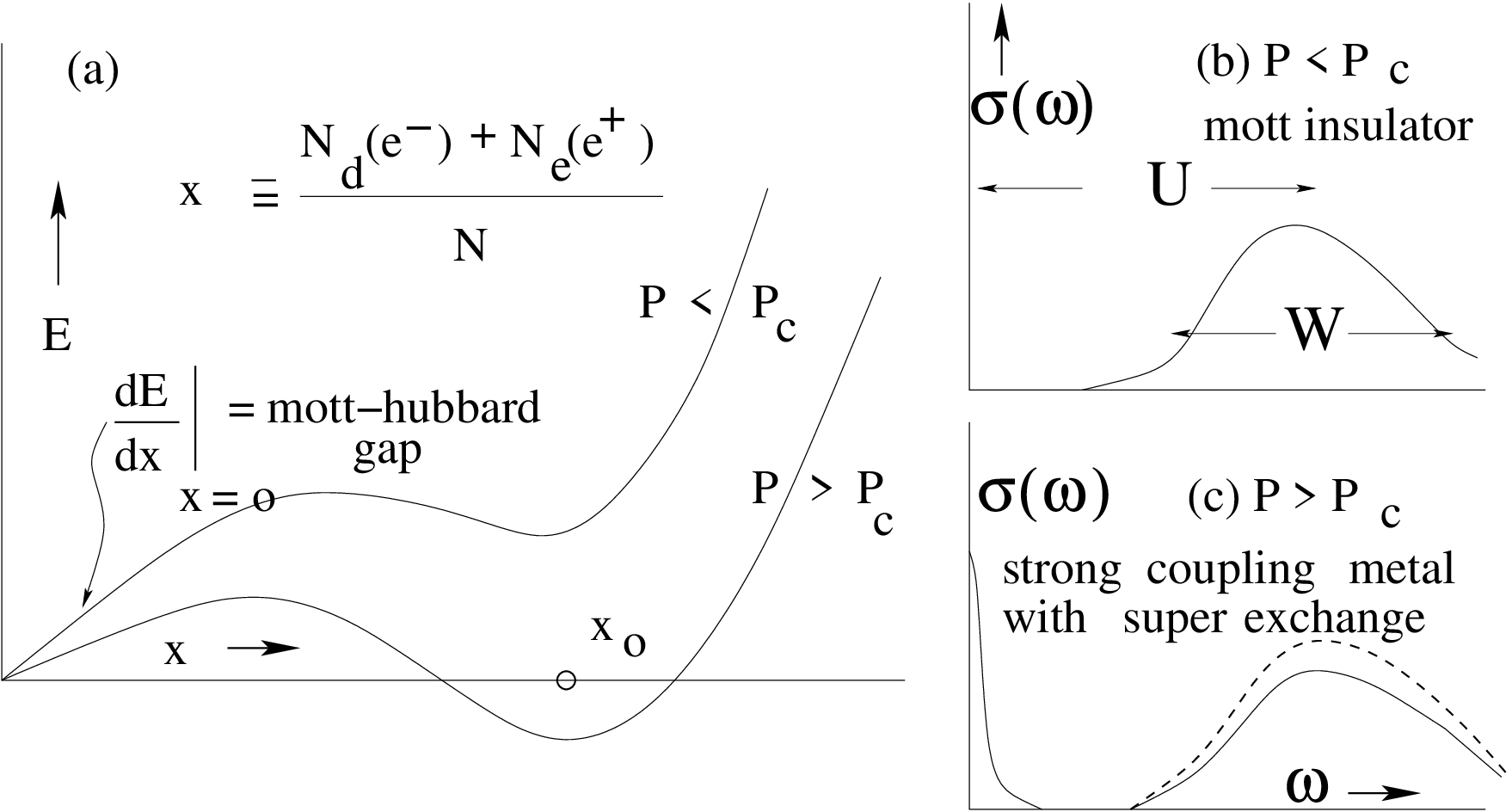}
\caption{\label{fig2} a) Energy of a half filled band above and below the
critical pressure $P_c$, as a function of
$x = \frac{N_d(e^-) + N_e(e^+)}{N}$. Here $N_d(e^-) = N_e(e^+)$ are
the number of doubly occupied ($e^-$) and number of empty sites ($e^+$);
total number of lattice sites N = total number of electrons.
Optimal carrier density $x_0 \equiv \frac{2N_0}{N}$ is determined by long
range part of coulomb interaction and superexchange energy.
b) and c) Schematic picture of the real part of the frequency
dependent conductivity on the insulating and metallic side close
to the Mott transition point in a real system. W is the band width.
}
\end{figure}
\begin{figure}
\includegraphics*[width=6.0cm]{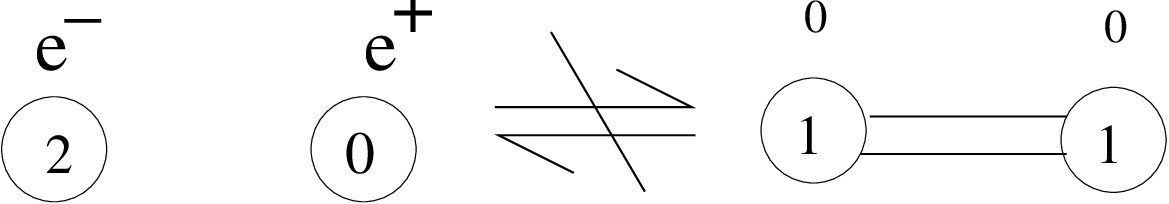}
\caption{\label{fig3} Forbidden hoping process,i.e., absence of annihilation of
$e^+$ and $e^-$
at low energies in our strong coupling metal. Double line represents
a spin singlet (valence) bond.
}
\end{figure}
\section*{\textsc{Superconductivity in Organic Solids}}

Superconductivity in organic molecular conductors is
a well developed field\cite{yamaji}. From a modest 1 K in Bechgard
salt, the superconducting Tc has increased to a value ~ 13 K in ET
salt family. This is remarkable, considering the low carrier
density, $n \sim 10^{20}/cm^3$in organics. Various ideas including
spin fluctuation mechanism of superconductivity has been discussed
to explain superconductivity in 2D organics. In my opinion they were
unsatisfactory.

One generic property of this system is that, after taking care of
crystallographic doubling of unit cell, it is well
described\cite{fukuyamaOSC, McKenzeeOSC} as half filled single band
system; i.e., one electro per Wannier orbital. Often these systems
exhibit Mott insulator to superconductivity transitions, either
under external pressure or chemical pressures.

What is the physics behind these Mott insulator to a superconductor
transition ? Firstly, it is a strong first order transition. A large
Mott gap (comparable to band width) collapses to zero value. Long
range coulomb interaction drives the transition first order. This is
missing in the standard Hubbard model; consequently it predicts a
continuous vanishing of the Mott Hubbard gap, across the
Mott-Hubbard transition. I observed\cite{GBosc} that, in
experiments, the optical conductivity $\sigma(\omega)$ retains the
upper Hubbard band feature nearly intact across the transition. The
only change is the emergence of a Drude peak at low frequencies; the
weight of the Drude peak indicates a small density of mobile
carriers (often as low as 5 to 10 \%).

Based on the above observation, I suggested that across the Mott
transition, the Mott insulator retains its integrity, in the sense
of survival of super exchange on the conducting side (figure 2). The only new
aspect is that a small and equal density of mobile positive and
negative charge carriers (doublons and holons) have been
spontaneously generated. These carrier density are individually
conserved (figure 3) and governed by the physics of long range coulomb
interaction. In other words, I gave a new interpretation of Mott
transition in theses systems as a process of self doping a Mott
insulator.

These suggestions implied immediately a close connection of the
mechanism of superconductivity to that in cuprates, where the doping
is external. I developed this idea further and introduced a 2
species t-J model and discussed how superconductivity arises there.

My conclusion\cite{GBosc} is that superconductivity in organics is
based on RVB mechanism. The new feature is that preexisting neutral
singlets get charged across the Mott transition and produce
superconductivity, through a process of self doping rather than
external doping. The superconducting Tc is determined by, apart from
other factors such as superexchange, the density of self doping. As
self doping increases beyond the optimal value (achieved by
increasing pressure) superconductivity quickly disappears, as seen in
the experiment. Recent theoretical works\cite{ZhangOSC, nandiniOSC,
MckenzeeOSC} essentially corroborate my view point, albeit with some
minor differences.

In a very recent work Kanoda\cite{kanodaRVB} group have reported
interesting results, in my opinion offering a direct support to RVB
physics in one of the members of the ET salt family in the Mott
insulating phase. They find evidence for a pseudo fermi surface like
excitations from magnetic and specific heat measurements. This is
likely to be a first example of realization of spinon pseudo fermi
surface in 2D, suggested by Anderson and realized in BZA theory.
This ET salt is a Mott insulator with enhanced near neighbour multi
spin couplings, in view of smaller Mott-Hubbard gap. These couplings
seem to frustrate antiferromagnetic order and really stabilise a
quantum spin liquid with a pseudo fermi surface for spinons.

\section*{\textsc{Boron Doped Diamond}}

Diamond is known to be one of the best insulators. It has a large
band gap of 5.6 eV. In a remarkable recent work Ekimov
and collaborators\cite{ekimovNature} have managed to convert diamond
to a superconductor by doping with boron; i.e., diamond:B. It is well known that small traces of boron impurities is responsible for the captivating blue colour of diamond; however, heavy doping makes it dark and superconducting! The superconducting Tc has steadily increased from about 4 K to nearly 12 K, with increasing doping and improved
material characteristics using MOCVD preparation methods\cite{bustarret, Takano}.

\begin{figure}
\includegraphics*[width=8.0cm]{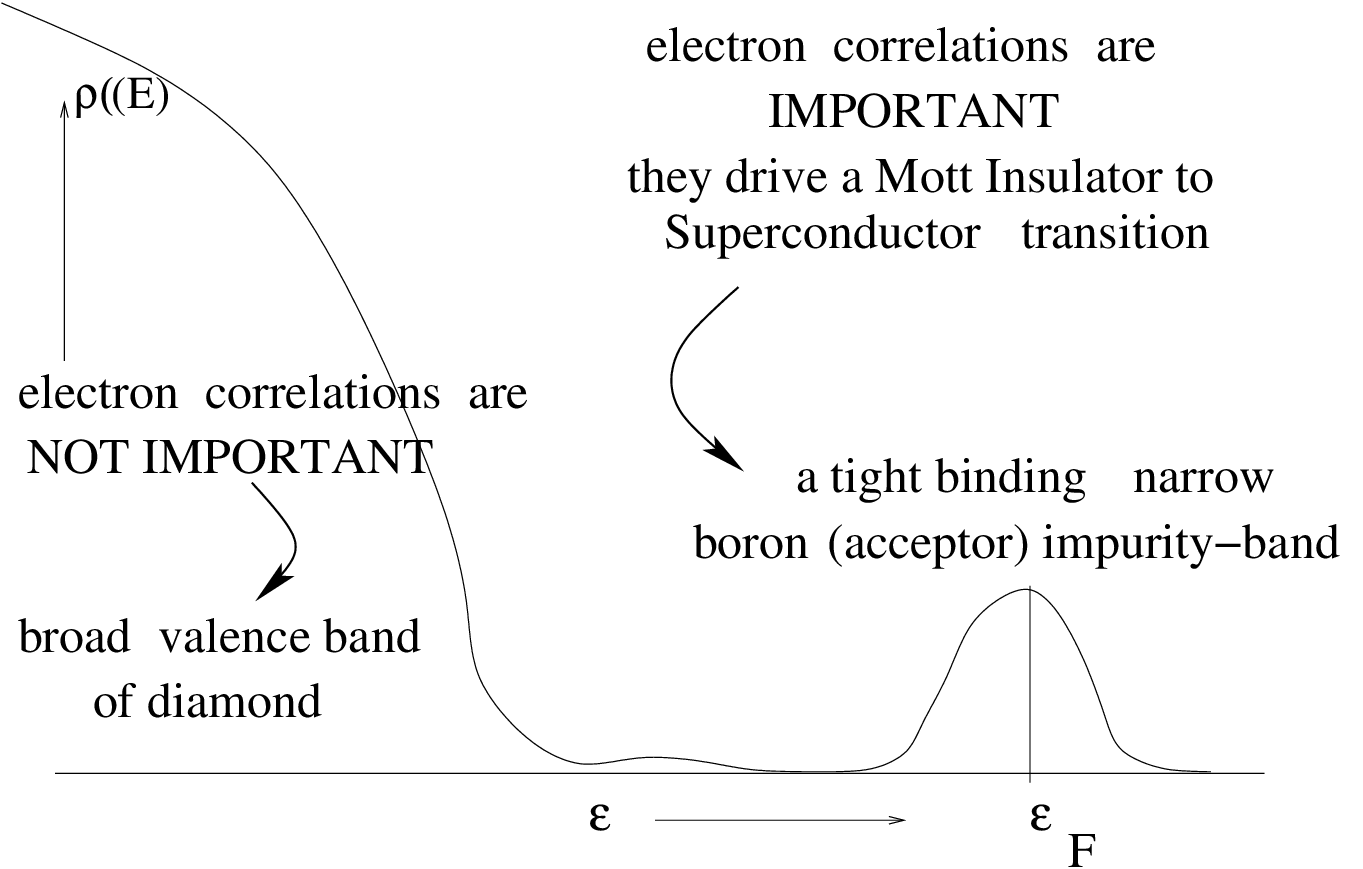}
\caption{\label{fig1}
Hole density of states (schematic) in boron doped diamond, an uncompensated
p-type semiconductor. Holes of acceptors form a strongly correlated and
impurity band at commensurate filling. Anderson-Mott insulator to
superconductor transition is suggested to take place in the impurity band
as we increase boron density (figure 5).}
\end{figure}
\begin{figure}
\includegraphics*[width=8.0cm]{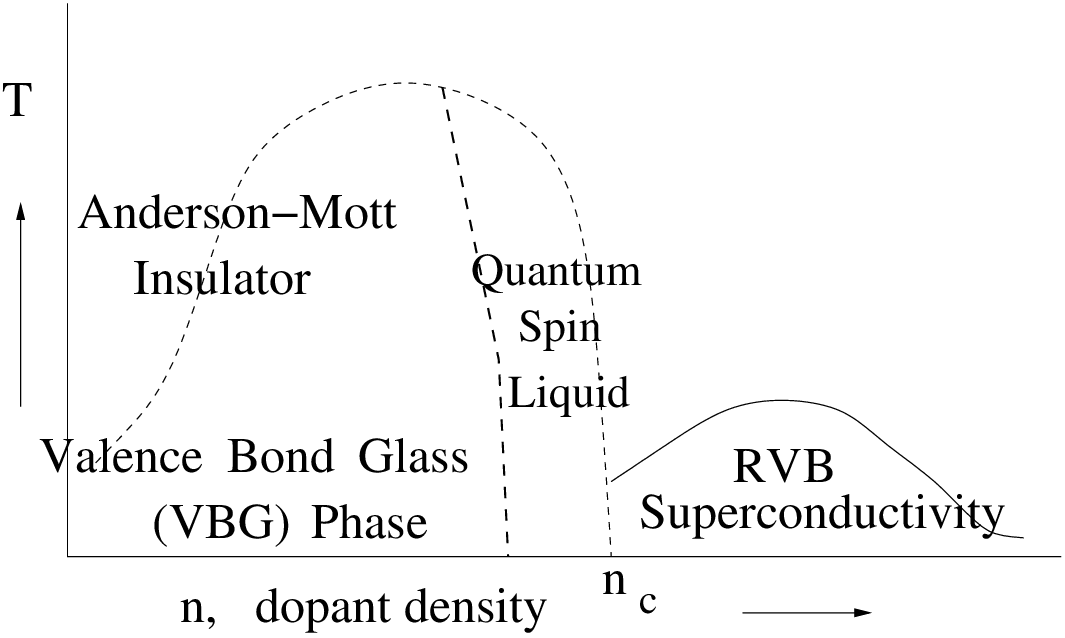}
\caption{\label{fig1} Schematic Phase Diagram as a function of
dopant density in Diamond:B, an uncompensated case.}
\end{figure}

 I have developed a theory\cite{GBdmnd}, based on phenomenological and
microscopic grounds that superconductivity takes place in the
impurity band introduced by boron substitution, across the insulator
to metal transition. Briefly, a substituted boron has a nice sp$^3$
tetrahedral bonding with neighbouring carbon atoms, except that there
is a missing electron, i.e. a hole. This hole resides in one of the
3 fold degenerate impurity states, at about 0.37 eV above the top of
the valence band. When the boron density is low, the holes are
localised in their respective hydrogenic type of impurity states and
well isolated. Holes, instead of getting delocalized into extended
states, remain in their home site, because of an effective U (hole
affinity - hole binding energy) $>$ impurity band width W; i.e., it
costs energy to ionise and delocalized a hole. It is a Mott insulator
formed of impurity states (figure 4, 5). The holes stay in their impurity states
and virtual fluctuations to neighbouring impurity sites leads to
antiferromagnetic superexchange interaction. This leads to spin
singlet coupling. Since the impurity states are randomly distributed
in space, a spin finds its closest neighbour and forms a valence
bond; this leads to valence bond solid (glass) phase, very similar
to what has been studied for the case of Si:P. The spin half
character of the hole in the impurity state and the orbital
degeneracies stabilise a valence band glass phase rather than a spin
glass phase.

As we increase boron concentration, we expect a Mott insulator to
metal transition, in the impurity band subsystem. Since we have an
uncompensated doping, randomness and Anderson localisation issues
are only secondary. We can imagine the impurity state subsystem as a
lattice of hydrogen atoms whose lattice parameter is decreased as
increasing dopant density. As we approach Anderson-Mott transition
point, the impurity state wave functions strongly overlap; i.e., the
inter impurity distance is comparable to the size of the impurity
state wave function (effective Bohr radius $a_B^*$). Valence bond
resonance increases and valence bond glass melts. We get a quantum
spin liquid in a disordered lattice (figure 5).

The resonating singlets are the preformed pairs. They are neutral.
Across the first order Mott transition, the Mott-Hubbard gap
collapses from a finite value to zero, by a process of self doping
of the Mott insulator. That is, the Mott insulator continues to be a
Mott insulator with valence bond resonance, except for a spontaneous
creation of a small density of delocalized $B^+$ and $B^-$ species.
This mechanism of superconductivity is very similar to our mechanism
for the organics, outlined in the previous section. In fact the
carrier density and the size of the molecular orbital in organics
and the impurity wave functions in diamond:B are similar in size
leading to a similar value of Tc.

In the literature, at least three different phonon mechanism, which
put the doped holes at the top of the valence band, in extended
states have been proposed\cite{dmndPhononMechanisms}. Even liberal
estimates of Tc give a value small compared to experiments. Various
phenomenology, particularly large value of low temperature intrinsic
resistivity and recent ARPES results\cite{yokoyaNature} indicate
that the carrier mean free path are comparable to nearest
boron-boron distance. There are also other experimental
evidence\cite{NakamuraXAS} for the existence of an impurity band in
the superconducting state, suggesting that the origin of short mean
free path of carriers is not necessarily due to randomness. It is
likely to be the effect of strong correlation within the impurity
band.

\section*{\textsc{Na$_x$CoO$_2$$\cdot$yH$_2$O, an Icy Superconductor}}

\begin{figure}
\includegraphics*[width=6.0cm]{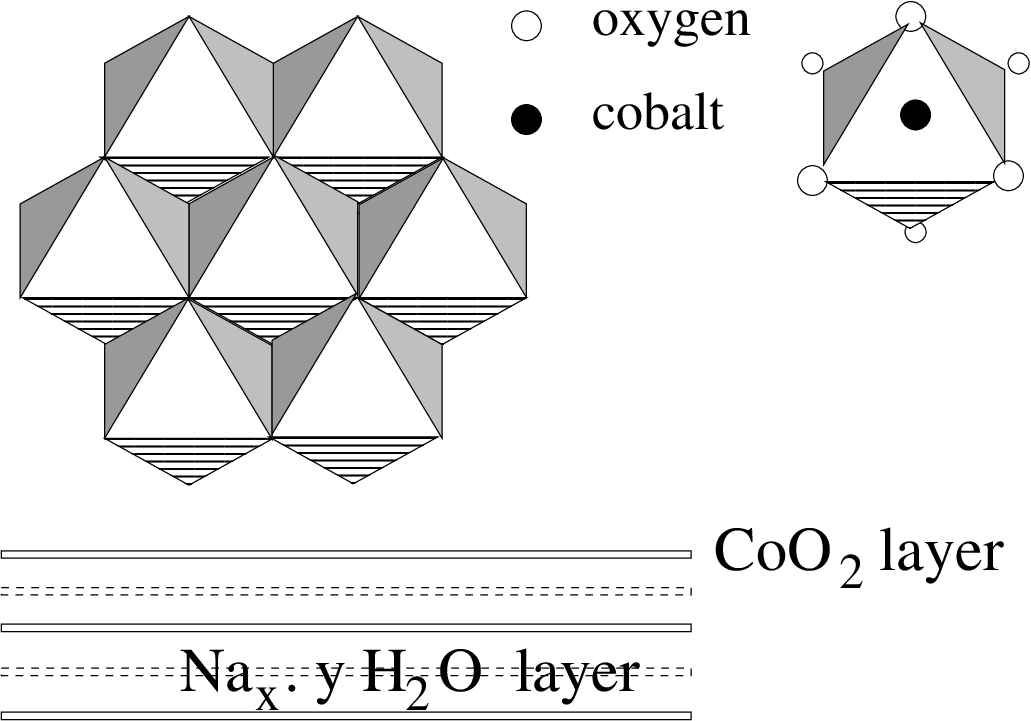}
\caption{\label{fig1}
A triangular network of edge sharing oxygen octahedra. Co atoms are
at the center of the oxygen octahedra. Each Na$_x$.yH$_2$O layer is
sandwitched by two CoO$_2$ layers.}
\end{figure}

\begin{figure}
\includegraphics*[width=6.0cm]{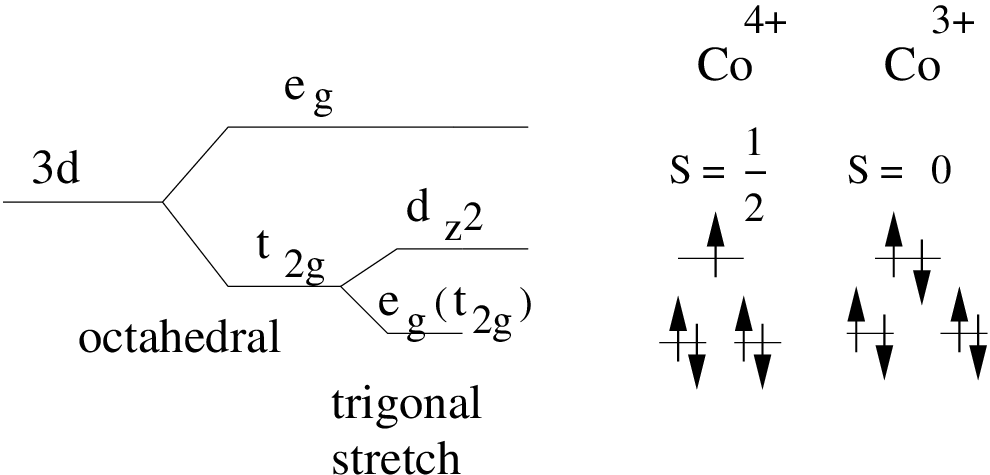}
\caption{\label{fig1} Crystal field split 3d levels of cobalt. }
\end{figure}

\begin{figure}
\includegraphics*[width=6.0cm]{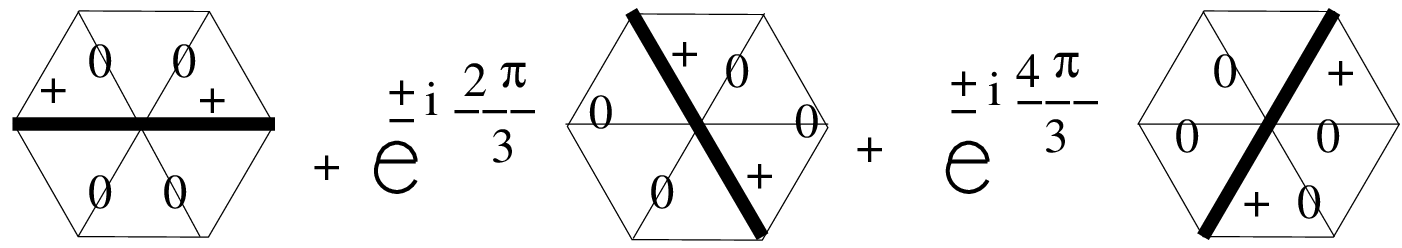}
\caption{\label{fig1} Relative phases of cooper pair amplitudes
($\Delta_{ij} \neq 0$ on dark bonds) in PT violating
$d_1 \pm id_2$ states.  }
\end{figure}

\begin{figure}
\includegraphics*[width=6.0cm]{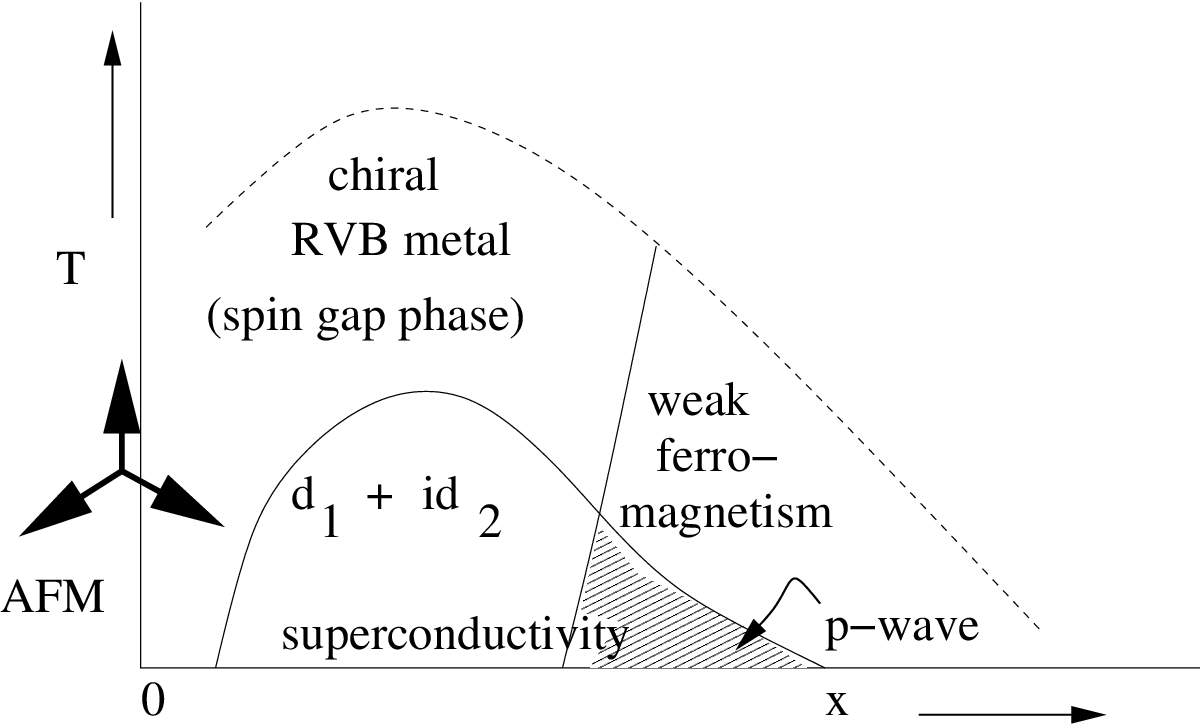}
\caption{\label{fig1}The schematic $x-T$ phase diagram. }
\end{figure}

New superconductors and novel materials continue to be discovered by
Japanse groups, thanks to their concerted efforts in materials
science with an eye not only on technology but also basic science.
Historically, many systems exhibiting RVB physics have been
discovered by the Japanese groups, including a Shastry Sutherland
compound SrCU$_2$ (BO$_3$)$_2$, alluded to earlier. Quickly following the footstep of discovery of
superconductivity in \mgb by Akimitsu group, a Tsukuba group
synthesized\cite{cobTsukuba} a layered $\rm{Na_xCoO_2}$ that becomes
superconducting only when it is intercalated with water: \nxcob
(figure 6).

Sitting in Chennai and I got a news of this icy superconductor by
email, through a superconductivity e-group. It became clear that it
is a doped spin half orbitally non-degenerate Mott insulator on a
triangular lattice (figure 7). It is indeed a long sought after,
doped spin-\hlf triangular lattice system ! Absence of orbital
degeneracy in \nxcob was my conjecture based on simple estimates:
that is, the small fermi surface pockets that appeared in David
Singh's electronic structure calculation\cite{davidSingh} should
infact disappear due to correlation effects, leaving a hole like
band around the $\Gamma$ point. It worked ! Later ARPES experiments
showed\cite{cobARPES} a single circular fermi surface, validating my
single band model hypothesis and a non standard sign of the hopping
integral.

I worked out an RVB theory quickly\cite{GBcob}. What was novel was,
the possibility of an important chiral RVB state as the reference
Mott insulator. In an earlier work Lee and Feng\cite{tkLee},
inspired by Kalmayer-Laughlin's chiral spin liquid
state\cite{KalmayerLaughlin}, had in fact found a PT violating RVB
mean field solution, where every triangular plaquette contained a
$\frac{\pi}{2}$ RVB magnetic flux. I showed that on doping, the
insulating chiral spin liquid will continue into a PT violating
chiral singlet superconductor, having a $d_{x^2-y^2} + id_{xy}$ (or
briefly, d+id) order parameter symmetry (figure 8).
 Subsequent
theoretical analysis by Kumar-Shastry\cite{ShastryKumar}, Dung-Hai
Lee, Patrick Lee and Wang\cite{dhLeeCOB} have supported the above
RVB scenerio. Within the RVB scenario, there is a possibility of
chiral p-wave superconductivity at high doping end (figure 9). There
are also other proposals of spin triplet
superconductivity\cite{Pwave}. On the experimental front, there is
an intense effort, using magnetic resonance studies, to find the
order parameter symmetry by Nagoya and Kyoto and other groups.
Recent results from Nagoya group\cite{sato2006}, confirm their
earlier findings and give strong evidence for spin singlet pairing.
The issue of gap is still not settled. Earlier $\mu$SR
studies\cite{muSRcob} did not see any parity violating orbital
magnetic field, making a PT violating state suspect. However, this
result should be carefully analyzed, because of a possible invasive
character of muon, through polarization of the H$_2$O dipoles. I
have suggested \cite{GBwaterCOB} that a local polarization of H$_2$O
molecules might destabilize superconductivity and stabilize a
competing charge order locally.

I had also suggested\cite{GBcob} that, in between the PT symmetric
metallic state and PT violating superconducting state, an
intermediate PT violating metallic (a chiral metal) phase should be
present , over a finite temperature interval (figure 9). Increase
sample quality should enable one to search for this PT violating
metallic state experimentally.

Heavily doped \ncob, has an interesting metallic state, which
exhibits a coherent charge transport, like a good metal; however
spins are incoherent as seen by a non-Pauli, Curie magnetic
susceptibility ! This phase has been called a Curie metal by the
Princeton group\cite{OngCurieMetal}. Combining the above with some
possible signatures of Luttinger volume anomaly seen in
ARPES\cite{cobARPES}, I have suggested\cite{GBdoublon} a phase
called `Quantum Charge Liquid'. This is a natural generalization of
RVB phase to heaavily doped Mott insulators.

\section*{\textsc{RVB and spin-1 Collective mode\\ in single Graphene Layer}}

Graphite was a play ground for RVB ideas in the hands of Pauling.
One should have expected some unique signature in low energy
physical properties from RVB physics. Surprisingly no one seems to
have looked for possible consequences of RVB phenomenon in graphite.
Historically, with the advancement of electronic structure
calculations and a variety of magnetic field dependent measurements,
such as de Haas van Alfven effect, the single electron theories have
been reigning supreme. One possible reason behind is that the subtle
RVB effects of a 2D graphene sheet, at low energy, are being masked
by the finite interlayer electron tunnelling matrix element
($t_{\bot} \sim 0.2 eV$), which gives rise to small cylindrical
fermi surfaces around the K and K' points in the BZ.

\begin{figure}
\includegraphics*[width=8.0cm]{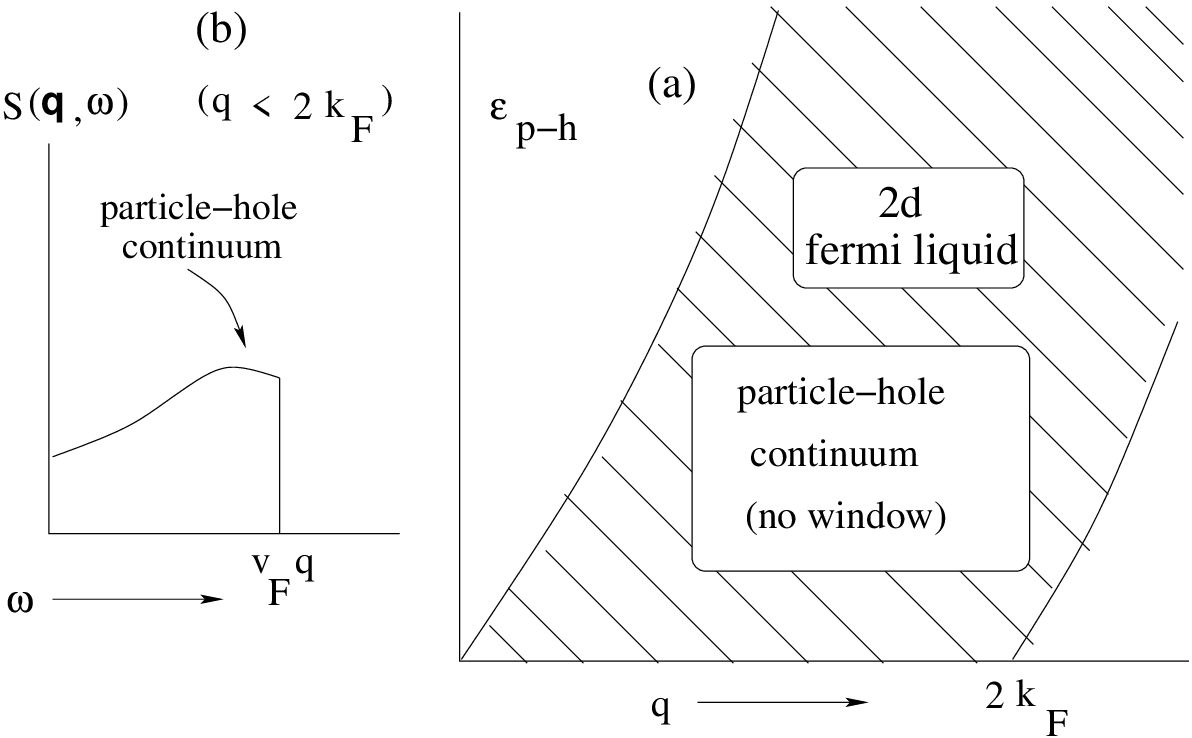}
\caption{\label{figxx} (a)particle-hole continuum without a `window' for
a 2d fermi gas. (b) $S({\bf q},\omega)$  for $q < 2k_F$.
}
\end{figure}

\begin{figure}
\includegraphics*[width=8.0cm]{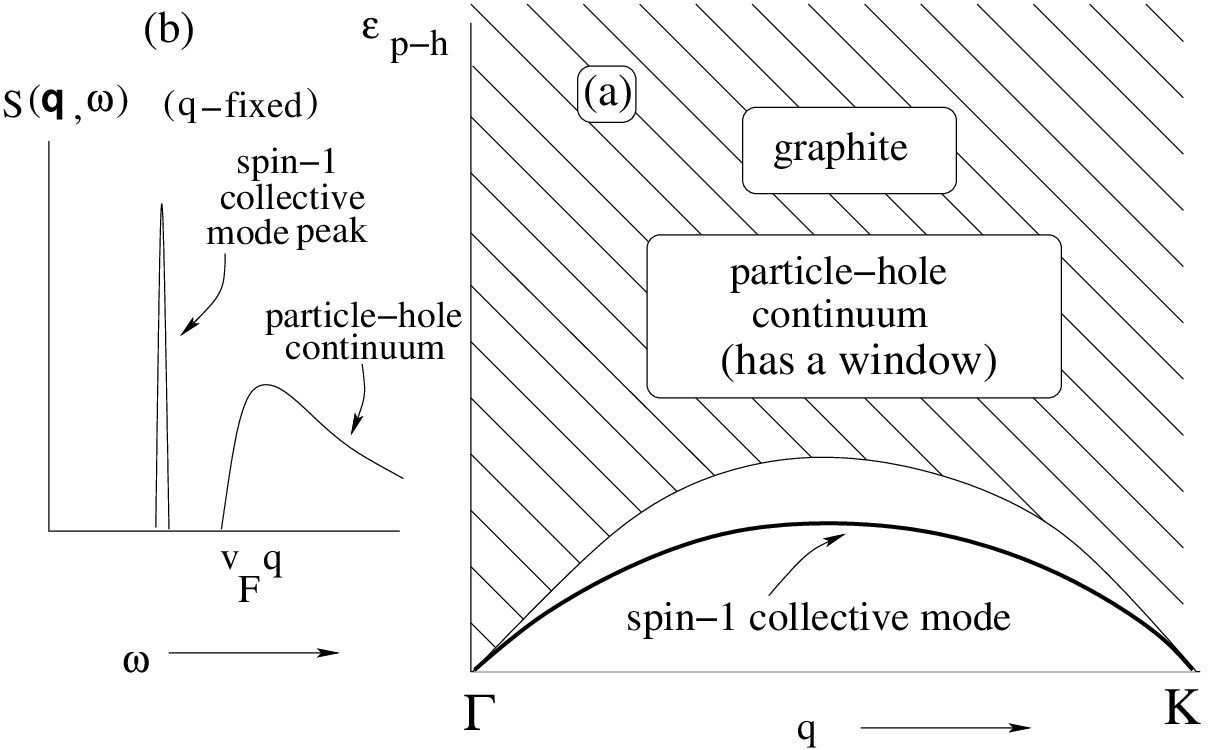}
\caption{\label{figxx} (a)particle-hole continuum with a `window' for
graphite. (b) $S({\bf q},\omega)$  for $q > q_c (\sim \frac{1}{50}
\frac{\pi}{a})$}
\end{figure}

In a recent paper\cite{GBakbar}, Akbar and I investigated effects of
electron electron interaction in a single graphene sheet, using a
simple Hubbard model. Graphene is a semi metal, where valence and
conduction bands meet at K and K' points in the BZ. That is the
fermi surface is shrunk to two points. Around these two points the
band structure locally resembles a Dirac cone. This leads to an
interesting gapless particle-hole continuum, that is very different
from the standard 2D particle-hole continuum (figure 10). Infact the
graphene particle-hole continuum (figure 11) has a big window. It
resembles particle-hole continuum of a 1D fermi gas rather than a 2D
fermi gas.

One effect of electron electron interaction is to modify the
excitation spectrum. We studied the particle-hole excitation
spectrum by a sraight forward RPA analysis, looking for spin-1
collective mode or triplet exciton. Based on what happens in
molecules such as benzene, anthracene etc., which are finite pieces
of graphene, we expected a spin-1 collective mode to emerge in some
region of the window. To our pleasant surprise we found that a
gapless spin-1 branch emerged in the full window, that is, all over
the BZ. Thus a new spin-1 collective mode branch has been predicted
for a single graphene sheet. Its energy ranged from zero to about 2
eV. In real graphite, inter layer coupling modifies the spectrum
somewhat,particularly below about 0.2 eV.

What is the spin-1 spectrum to do with RVB ? If electrons in a
graphene sheet are non interacting, singlet correlation exists in
the ground state only because of kinematics imposed by Pauli
principle. That is, a Bloch state is occupied by two electrons with
opposite spin, to make a spin singlet in k-space. This minimal
singlet correlation is kinematic in origin. However, when one
introduces finite U, repeated collisions in the spin singlet channel
enforce spin singlet correlations in the ground state. If U were
large compared to the band width, we would have had a Mott insulator
and these collision processes would have been called superexchange
processes. But graphene is not a Mott insulator. Still some kind of
kinetic exchange processes continue to exist, which give an enhanced
near neighbor singlet correlations in the ground state, compared to
the free fermi gas.

In other words, the emergence of spin-1 collective mode indicates a
coherent modification of the free fermi gas state, into an RVB state
or a quantum spin liquid state. If there is an RVB physics and it it
is a metal, why a finite temperature superconductivity is absent in
graphite ? I have found, in a recent work\cite{gbMgB2}
that the development of RVB correlation (pre existing singlet pairs)
in a graphene sheet fail to make it a superconductor, because of
vanishing of the single particle density of states at the fermi
level.

Recently, single graphene states have been isolated and studied,
from quantum Hall effect point of view\cite{grapheneQHE}, yielding
spectacular integer quantization of Hall conductance. It will be
interesting to study single graphene sheet and look for consequences
of RVB correlations, including large superconducting fluctuations,
spin-1 collective modes and its effects.

It is a pleasure to recall important contributions from the
Chanchal Majumdar group on quantm theory of magnetism, including
the Majumdar-Ghosh model, that is being celebrated in this meeting.

\end{document}